\documentclass[12pt]{article}

\usepackage{lineno}

\usepackage[margin=1in]{geometry}
\usepackage{amssymb,amsmath,latexsym}                        
\usepackage{graphicx}
\usepackage{cite}
\usepackage{xcolor}
\usepackage{float}

\usepackage{amsthm}
\usepackage{multicol}
\usepackage{authblk}
\usepackage{subfigure}

\usepackage[colorlinks]{hyperref}

\usepackage{adjustbox}
\usepackage{bbm}
\usepackage{mathtools}

\usepackage[titletoc]{appendix}
\date{\today} 

\begin{document}

\title{Ultimate position resolution of pixel clusters with binary readout for particle tracking}
\author[1,2]{Fuyue Wang}
\author[2]{Benjamin Nachman}
\author[2]{Maurice Garcia-Sciveres}
\affil[1]{\normalsize\it Department of Engineering Physics, Tsinghua University, Key Laboratory of Particle and Radiation Imaging, Ministry of Education, Beijing 100084, China}
\affil[2]{\normalsize\it Physics Division, Lawrence Berkeley National Laboratory, Berkeley, CA 94720, USA}

\maketitle

\begin{abstract}
 Silicon tracking detectors can record the charge in each channel (analog or digital) or have only binary readout (hit or no hit). While there is significant literature on the position resolution obtained from interpolation of charge measurements, a comprehensive study of the resolution obtainable with binary readout is lacking. It is commonly assumed that the binary resolution is $\text{pitch}/\sqrt{12}$, but this is generally a worst case upper limit. In this paper we study, using simulation,  the best achievable resolution for minimum ionizing particles in binary readout pixels.  A wide range of incident angles and pixel sizes are simulated with a standalone code, using the Bichsel model for charge deposition. The results show how the resolution depends on angles and sensor geometry. Until the pixel pitch becomes so small as to be comparable to the distance between energy deposits in silicon, the resolution is always better, and in some cases much better, than  $\text{pitch}/\sqrt{12}$.
\end{abstract}

\section{Introduction}
\label{sec:intro}

\begin{figure}
\centering
\subfigure[]{\includegraphics[width=0.48\textwidth]{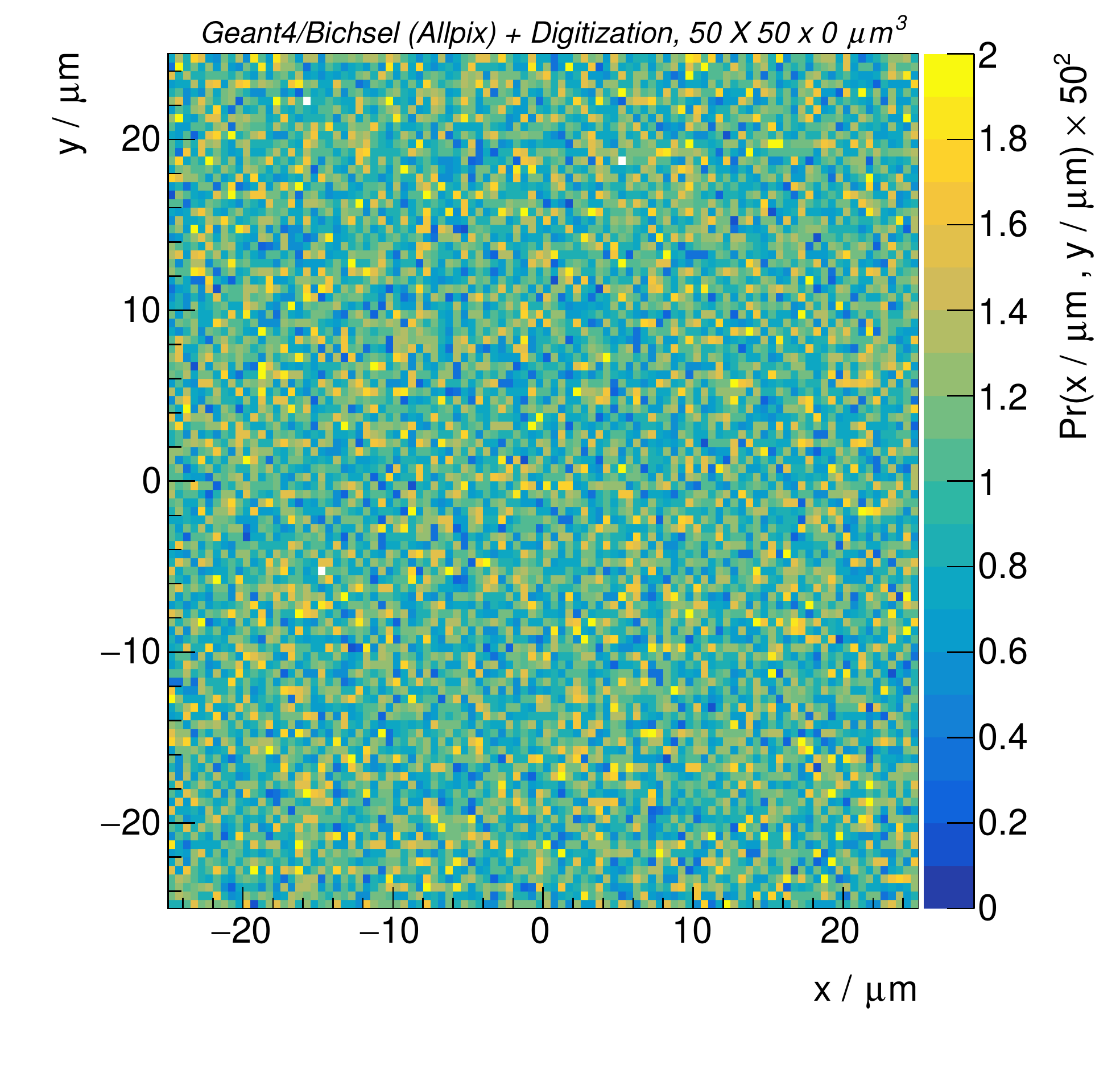}}
\hspace{0.1in}
\subfigure[]{\includegraphics[width=0.48\textwidth]{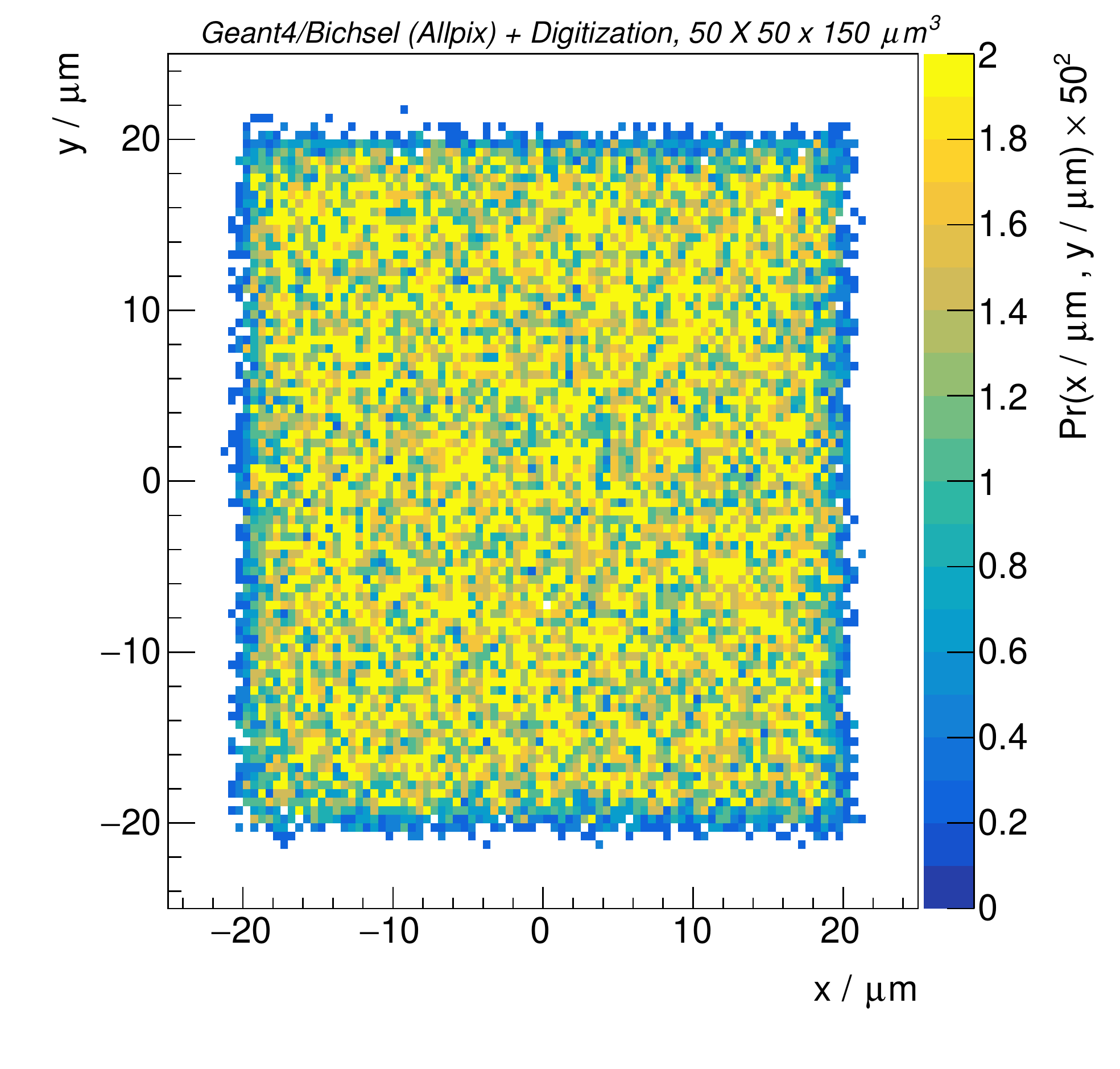}}
\caption{\label{fig:1hit}. Distribution of track positions in a single pixel cluster: 
(a) commonly assumed, which is only correct for an infinitely thin sensor; 
(b) actual for normal incidence tracks in a $50\times50\times150$ $\mu$m$^3$ sensor. }
\end{figure}

The spatial resolution of silicon strip and pixel detectors has been analyzed in detail before (see for example \cite{Gluckstern:1963,Turchetta:1993vu,Boronat:2014yya}). This prior work has focused on the resolution that can be obtained by interpolation of charge measurements in adjacent channels, on the charge deposition and transport processes, on signal to noise of the charge measurement, and on functional forms to calculate position from measured charges. However, the resolution limits in the case of binary readout (no charge information, just hit or no hit above a preset threshold) have not been fully explored. For example, it is commonly stated that a single channel hit has a spatial resolution of pitch/$\sqrt{12}$ -- the standard deviation of a uniform random variable on the interval $[0,\text{pitch}]$, but this is actually a worst case upper limit. Worst case means that no additional knowledge about the hit has been used, such as the observed cluster distributions in the detector that the hit belongs to, or the approximate incidence angles of the particle track producing the hit. Yet both of these things are known in practical applications, when fitting a track to a collection of hits. For example, Fig.~\ref{fig:1hit}(a) shows the assumed hypothetical distribution of true track position in a pixel that leads to the pitch/$\sqrt{12}$ result, compared to the actual distribution in Fig.~\ref{fig:1hit}(b) which shows the track positions for the case of single hit clusters in a $50\times50\times150$ $\mu$m$^3$ pixel sensor being crossed at normal incidence\footnote{We present all results in the absence of a magnetic field. If a magnetic field is present the incidence angles relative to the sensor plane must be replaced by angles relative to the charge drift direction. The equivalent of ``normal incidence" would thus be ``parallel to the drift direction".}. The RMS of the actual distribution shown is $0.78 \times \text{pitch}/\sqrt{12}$. The reason is that tracks near the edge of the pixel will produce 2-pixel clusters instead of 1-pixel clusters.  This charge sharing is related to the sensor thickness: in an infinitely thin sensor the track positions for 1-hit clusters would 
actually look like Fig.~\ref{fig:1hit}(a).
The beneficial effect of knowing the cluster size distribution becomes greater for larger clusters, as will be seen later. 

\begin{figure}
\centering
\includegraphics[width=0.7\textwidth]{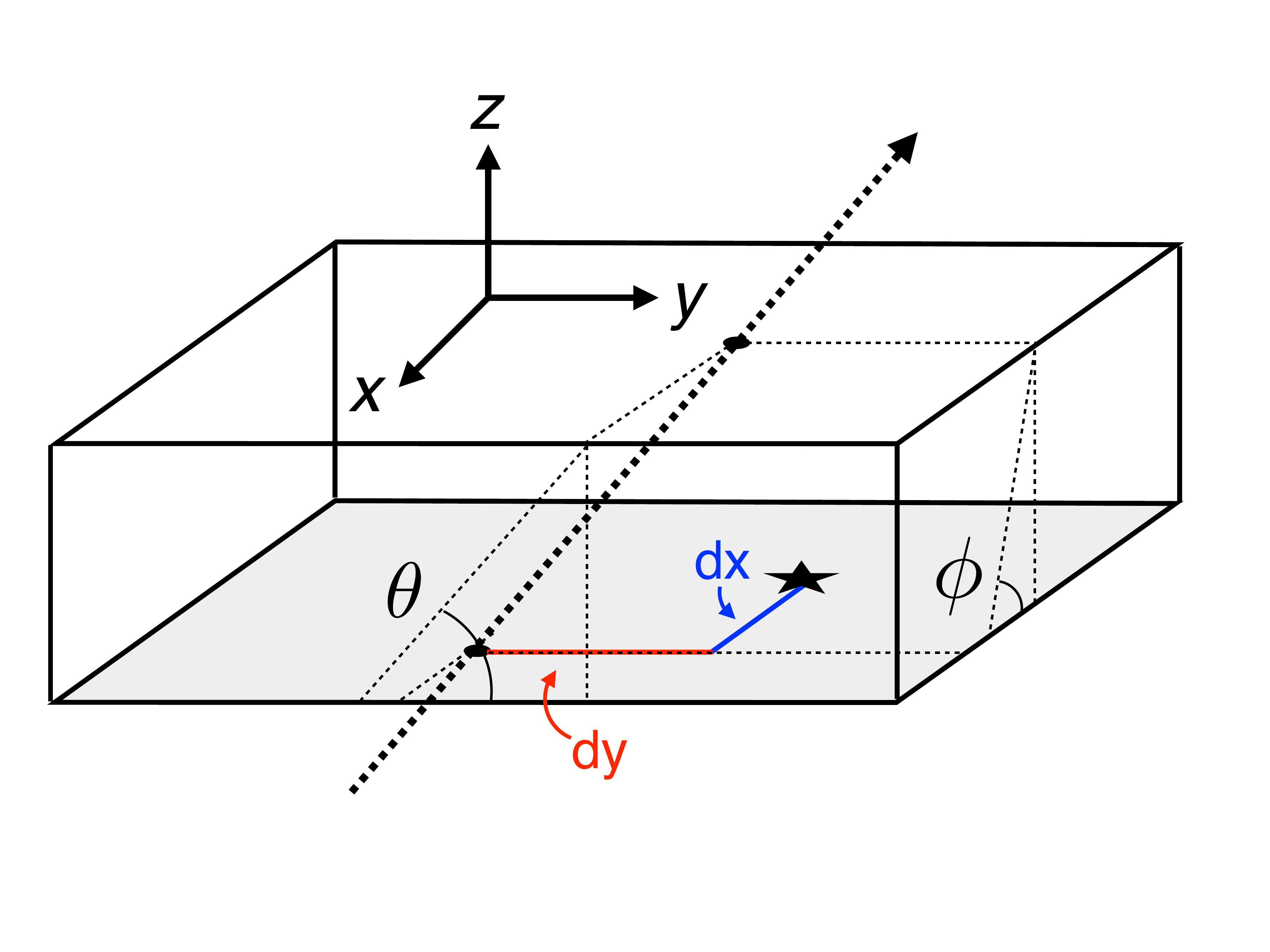}
\caption{\label{fig:residuals}Definition of coordinates and incidence angles and depiction of the residual 
on the sensor surface of incidence. The particle trajectory is shown as the heavy dotted line, with the sensor entrance  
and exit points are marked by ovals.  The estimated entrance point is indicated by a star. The 
separation between estimated and true entrance points on the 
plane of the sensor surface is decomposed into d$x$ and d$y$.} 
\end{figure}

The goal of this paper is to survey the position resolution of binary readout pixels for a variety of pixel sizes of current and future interest, as shown in Table \ref{tab:pitch}, covering the range of incidence angles found in practical applications.
The incidence angle is decomposed along two directions: polar ($\theta$) and azimuthal ($\phi$) as shown in Fig.~\ref{fig:residuals}. 
Colliding beam detectors contain a central {\em barrel} section of cylindrical geometry with axis along the colliding beams, 
while fixed target as well as forward elements of colliding beam detectors use sensor planes approximately perpendicular to the particle flux. 
In the barrel geometry the azimuthal range is small ($\phi<30^{\circ}$) and the polar range is larger ($\theta_{\rm max}>45^{\circ}$),
while in forward planes both angles are small. 
Hadron colliders use the variable pseudorapidity $\eta \equiv -\ln(\tan(\theta/2))$ to describe the polar angle. 
In this study we do not simulate a particular detector geometry, but rather a single sensor in a flux of incident 
particles, spanning the angular ranges $0<\phi<50^{\circ}$ and $0<|\eta|<2$. 

We simulate rectangular as well as square pixels, as the former are often used in barrel detectors, with the long dimension of the pixel along the beam direction, which is done in order to improve position resolution in the 
azimuthal direction without increasing channel density. The details of our simulation are presented in Sec.~\ref{sec:simulation}.
As we are exploring the limits of resolution, we do not include noise or electronic charge sharing (crosstalk) in the simulation. 
The impacts of noise and crosstalk, which are unavoidable in a real system, are assessed in Sec.~\ref{sec:sharing}.


\section{Simulation}
\label{sec:simulation}
The simulation is based on the Geant4 package~\cite{Agostinelli:2002hh}. We are concerned with the intrinsic resolution for minimum ionizing charged particles and therefore simulate a monochromatic source of 20\,GeV muons. For low momentum particles, the hit distance from the fitted track will be dominated by multiple scattering, and therefore this analysis is mainly of interest to relatively high momentum particles.   
Since the incidence angle will not be perfectly known when fitting a collection of hits, 
we uniformly smear the entrance angles in each simulation by 1\,mrad. In terms of multiple scattering of 20\,GeV
muons, the RMS scattering angle when traversing 3\% of a radiation length is about 0.1\,mrad, which shows that the applied smearing is indeed coarse~\cite{Patrignani:2016xqp}.
To ensure uniform illumination over at least one pixel, the source position is smeared uniformly by the pixel dimensions. 


We simulate a single sensor at a time illuminated by the above source, with a variety of sensor thicknesses and pixel sizes as shown in Table~\ref{tab:pitch}.  All simulations use a planar sensor geometry. We simulate the collection of electrons. 
The Bichsel model\cite{Bichsel:1988if} is used to predict the energy loss along the particle trajectory, and we convert the energy loss to 
charge via $1e^- = 3.6$ eV\cite{Agashe:2014kda}. 
For computational convenience we group electrons into $\mathcal{O}(10)$ clumps per pixel pitch, except for the smallest pixels simulated, for which the number of electrons per pixel pitch is already of order 10 and 
no clumping is needed (each electron is its own clump). 
Each clump is transported to the sensor surface assuming drift along the electric field plus diffusion perpendicular to the 
field equal to 2.5\,$\mu\text{m} \times \sqrt{d/300\,\mu\rm{m}}$\cite{Agashe:2014kda}, where $d$ is the drift length. 
The values chosen correspond to the diffusion length expected at $-10^{\rm \circ}$C with 1\,V/$\mu$m field. We assume a uniform drift field of 1\,V/$\mu$m
in all cases, which is above unirradiated depletion voltage for all thicknesses considered and is also consistent with 
saturation drift velocity~\cite{JACOBONI197777}.  Other sources of charge sharing such as electronic noise and capacitive coupling between readout chips are not considered, because the impact is checked to be small (see Sec.~\ref{sec:sharing}).

The total charge arriving at each pixel is then compared to a threshold and the pixel is considered hit if the charge is above 
threshold, and not hit otherwise. We use a threshold value of 1000\,e$^-$ for 150\,$\mu$m thick sensors and scale it linearly with 
sensor thickness.  

\begin{table}
\centering
\small
\begin{tabular}{ | c | c | }
\hline
\textbf{Pixel dimension ($x \times y$)} & \textbf{Sensor Thickness} \\
\hline
50\,$\mu$m $\times$ 250\,$\mu$m & 200\,$\mu$m  \\
\hline
50\,$\mu$m $\times$ 50\,$\mu$m & 150\,$\mu$m, 100\,$\mu$m \\
\hline
25\,$\mu$m $\times$ 100\,$\mu$m & 150\,$\mu$m, 100\,$\mu$m \\
\hline
25\,$\mu$m $\times$ 25\,$\mu$m & 100\,$\mu$m, 50\,$\mu$m  \\
\hline
10\,$\mu$m $\times$ 10\,$\mu$m & 20\,$\mu$m  \\
\hline
5\,$\mu$m $\times$ 5\,$\mu$m & 10\,$\mu$m  \\
\hline
2\,$\mu$m $\times$ 2\,$\mu$m & 10\,$\mu$m, 5\,$\mu$m  \\
\hline
\end{tabular}
\caption{Simulated pixel sensor geometries.}
\label{tab:pitch}
\end{table}

\clearpage

\section{Shape Classification and RMS Calculation}
\label{sec:calculation}

\begin{figure}[h!]
\centering
\subfigure[]{\includegraphics[width=0.49\textwidth]{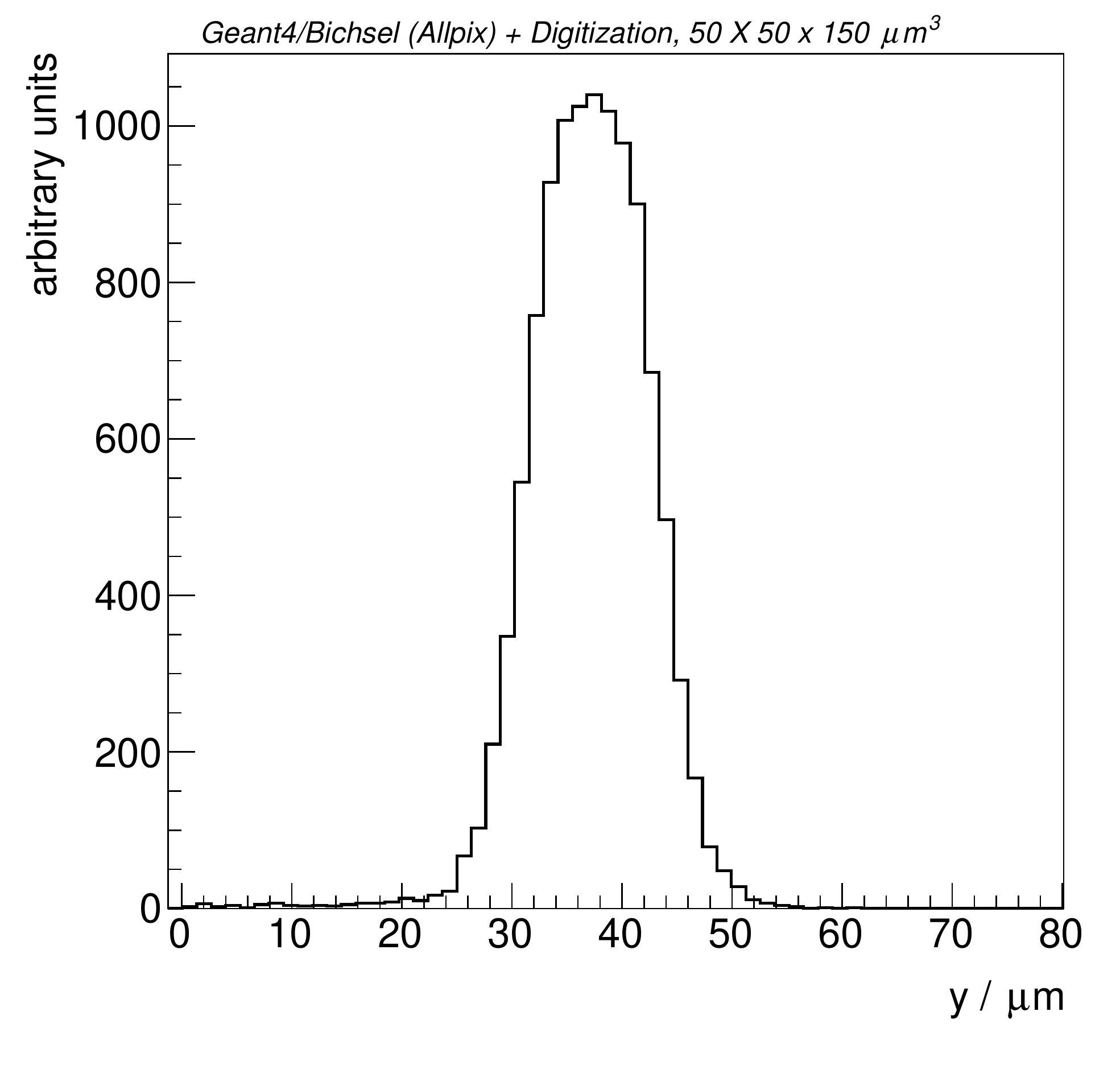}}
\subfigure[]{\includegraphics[width=0.49\textwidth]{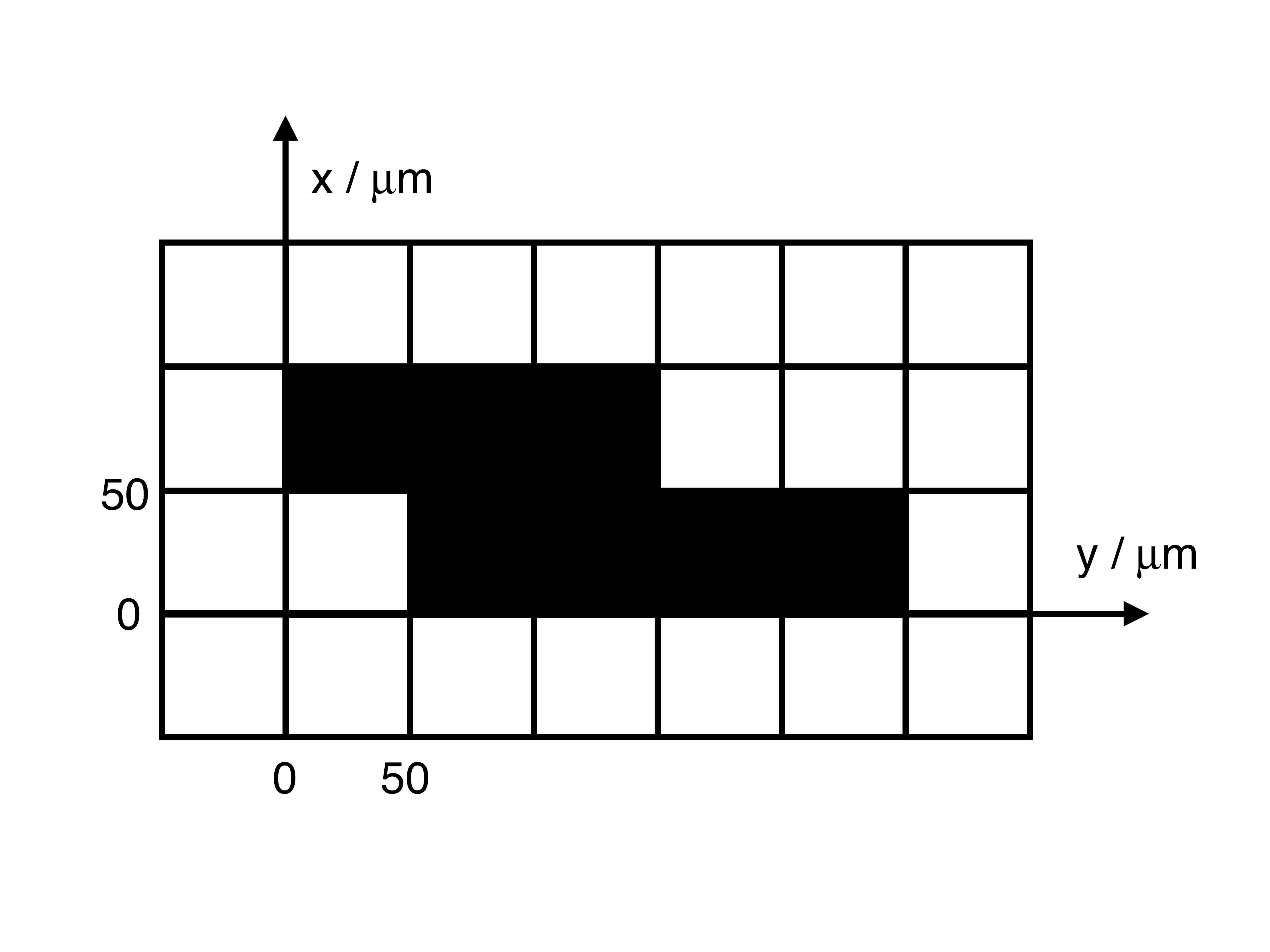}}
\caption{\label{fig:histexample} A histogram of entrance point $y$-coordinate (a) from all the clusters with the shape shown in (b). The pixel pitch is $50\times50\times150$ ${\mu}\text{m}^3$. The mean of this distribution gives the optimal estimator of position, while the RMS width gives the resolution. The incidence anglesare $\phi=10^{\rm \circ}$ and $|\eta|=1.0$}
\end{figure}

For each incidence angle we observe several distinct cluster shapes and for each shape we record the distribution of particle track entrance points to the sensor.  The optimal position estimator of each module given the shape and incidence angle is the mean of the entrance point distribution (the estimator that minimizes the mean squared error relative to the true position). Symbolically, the optimal position resolution is described by

\begin{align}
\label{eq:optimalres}
\sigma^2(x) \equiv \min\hspace{1mm} (d\text{x})^2 = \sum_\text{shapes $s$} \Pr(s)\min\hspace{1mm}(d\text{x}_s)^2= \sum_\text{shapes $s$} \Pr(s) \langle (x_\text{true}-\langle x_\text{true}|s\rangle)^2 \rangle ,
\end{align}

\noindent where $d\text{x}$ is defined in Fig.~\ref{fig:residuals}, $\Pr(s)$ is the probability of a particular shape s, $d\text{x}_s$ is conditioned on this shape $s$, and $\langle x_\text{true}|s\rangle$ is the average true entrance position given the shape $s$.  An analogous formula describes the optimal position resolution in the $y$ direction.  As a consequence of Eq.~\ref{eq:optimalres}, all that is needed to compute the optimal resolution is the RMS of the true position for all possible shapes.   Since our pixel information is binary, two clusters with the same hit pixel pattern are indistinguishable. We histogram the true entrance point independently for each cluster shape and compute the mean and RMS of the distribution. The mean is used as the optimal estimator for the position. As an example, Fig.~\ref{fig:histexample}\,(a) shows the true entrance point y coordinate histogram for the cluster shape shown in (b).  All position resolution results are shown normalized to the pitch/$\sqrt{12}$ assumption.

Enumeration of all possible cluster shapes is not practical. 
We therefore do not worry in advance about how many different shapes will be generated. 
All simulated events are processed and we simply create a new entrance point histogram for each new cluster shape
that appears. The procedure is as follows:
\begin{enumerate}
\item Simulate a new event.
\item Check if an identical cluster shape has already occurred by comparing the cluster to all existing histogram labels:
\begin{enumerate}
\item If the shape matches an existing label, add entrance point to that existing histogram;
\item If there is no match, create new histogram, label it with the cluster shape, and add the entrance point to this histogram.  
\end{enumerate}
\end{enumerate}
Two clusters are considered identical if they can be translated along $x$ and $y$ to lie exactly on top of each other, 
matching pixel-by-pixel. We do not consider rotations or reflections to produce a match. 
Furthermore, if $\Pr(s)$ is small, then the contribution to Eq.~\ref{eq:optimalres} is negligible.  
We therefore only include in the calculation of $\sigma^2$ those histograms with at least 10 entries, 
which means this shape occurs at least 10 times in the  simulation sample ($\Pr(s)\approx 0.01\%$). The fraction of events discarded by ignoring such very low probability shapes is about 3\%.
The number of included cluster shapes can be seen in Table~\ref{tab:Nbofshape}.


\begin{table}[h!]
\centering
\small
\begin{tabular}{ | c | c | c | c | c | c | c | }
\hline
& \textbf{$\phi=0$} & \textbf{$\phi=10^{\rm \circ}$} & \textbf{$\phi=20^{\rm \circ}$} & \textbf{$\phi=30$} & \textbf{$\phi=40^{\rm \circ}$} & \textbf{$\phi=50^{\rm \circ}$} \\
\hline
\textbf{$\eta=0$} & 8 & 8&9 &16 &23 &33 \\
\hline
\textbf{$\eta=0.4$} & 11&13 &12 &18 &26 &34\\
\hline
\textbf{$\eta=0.8$} & 26&18 &22 &27 &39 &43\\
\hline
\textbf{$\eta=1.2$} &50 &33 &34 &43 &54 &73\\
\hline
\textbf{$\eta=1.6$} &108 &42 &45 &51 &64 &121\\
\hline
\textbf{$\eta=2.0$} &104 &68 &71 &92 &109 &187\\
\hline
\end{tabular}
\caption{Number of cluster shapes occurring at least 10 times in our simulation for given incidence angles, 
for the $50\times50\times150$ ${\mu}\text{m}^3$ sensor. }
\label{tab:Nbofshape}
\end{table}

\clearpage

\section{Results}
\label{sec:results}
\begin{figure*}[h!]
	\centering
	\includegraphics[width=0.49\textwidth]{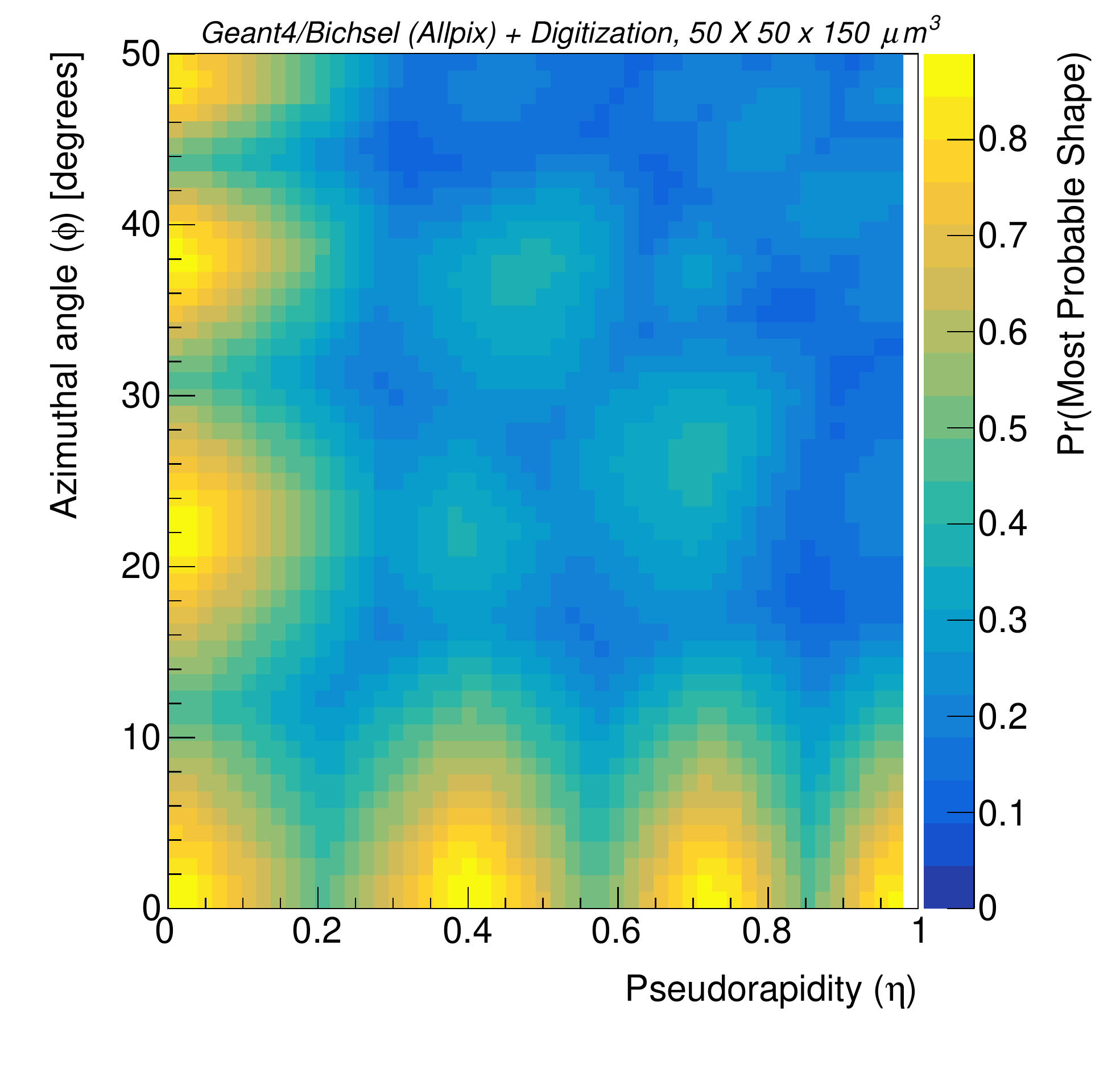}\includegraphics[width=0.49\textwidth]{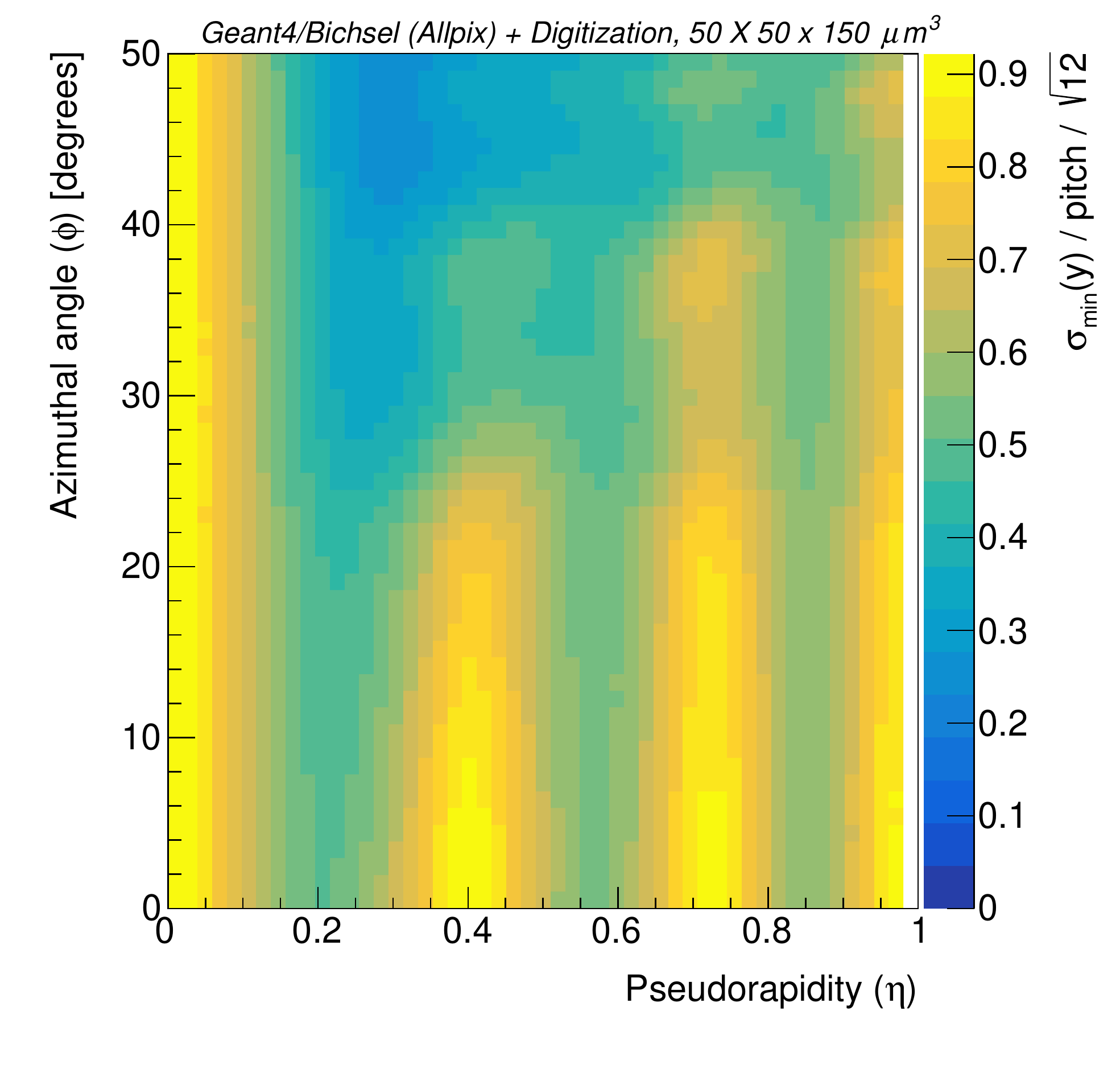}
	\caption{Left: The probability of the most frequent cluster shape for each $\phi$ and $\eta$ angle combination (maximum $\Pr(s)$).  Right: The optimal $y$ position resolution ($\sigma(y)$). }
	\label{fig:2dprob}
\end{figure*}

The position resolution of the binary readout depends on the variety of cluster shapes, which in turn depends on 
the incidence angle. Put another way, the resolution should be best when the maximum value of $\Pr(s)$ is small.  
This relationship is quantified by Fig.~\ref{fig:2dprob}, which shows the probability of that most popular shape and the optimal position resolution in the plane of $\phi$ vs. $\eta$ incidence angles, for a $50\times 50\times 150$ $\mu\text{m}^3$ pixel sensor.  As expected, regions of high (low) probability for the most probable shape correspond to a worse (better) position resolution and vice versa.  Since Fig.~\ref{fig:2dprob}(left) is nearly symmetric about $\phi=\eta$ in this angular range, the x position resolution plot is similar to Fig.~\ref{fig:2dprob}(right) with the $\phi$ and $\eta$ axis labels swapped. 

\begin{figure*}
	\centering
	\includegraphics[width=0.6\textwidth]{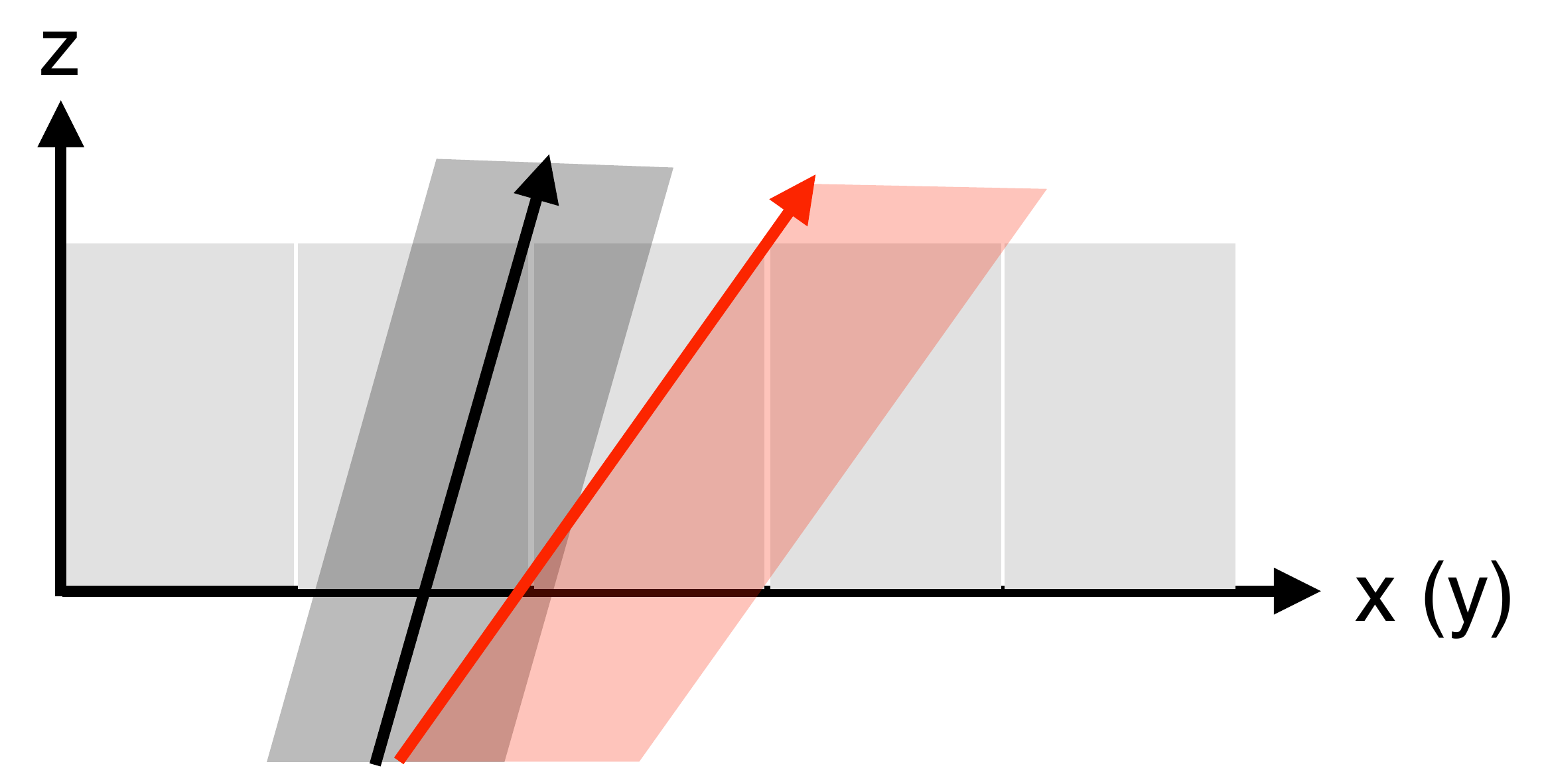}
	\caption{A side view of the pixel detector with arrows corresponding to two possible MIP trajectories.  The shaded red and black regions correspond to horizontal translations of the arrows.  The incidence angle of the black arrow is shallow enough to traverse either one or two pixels depending on its horizontal position.  However, the red arrow will always traverse two pixels, no matter how it is translated horizontally.}
	\label{fig:length}
\end{figure*}

\begin{figure*}
	\centering
	\includegraphics[width=0.5\textwidth]{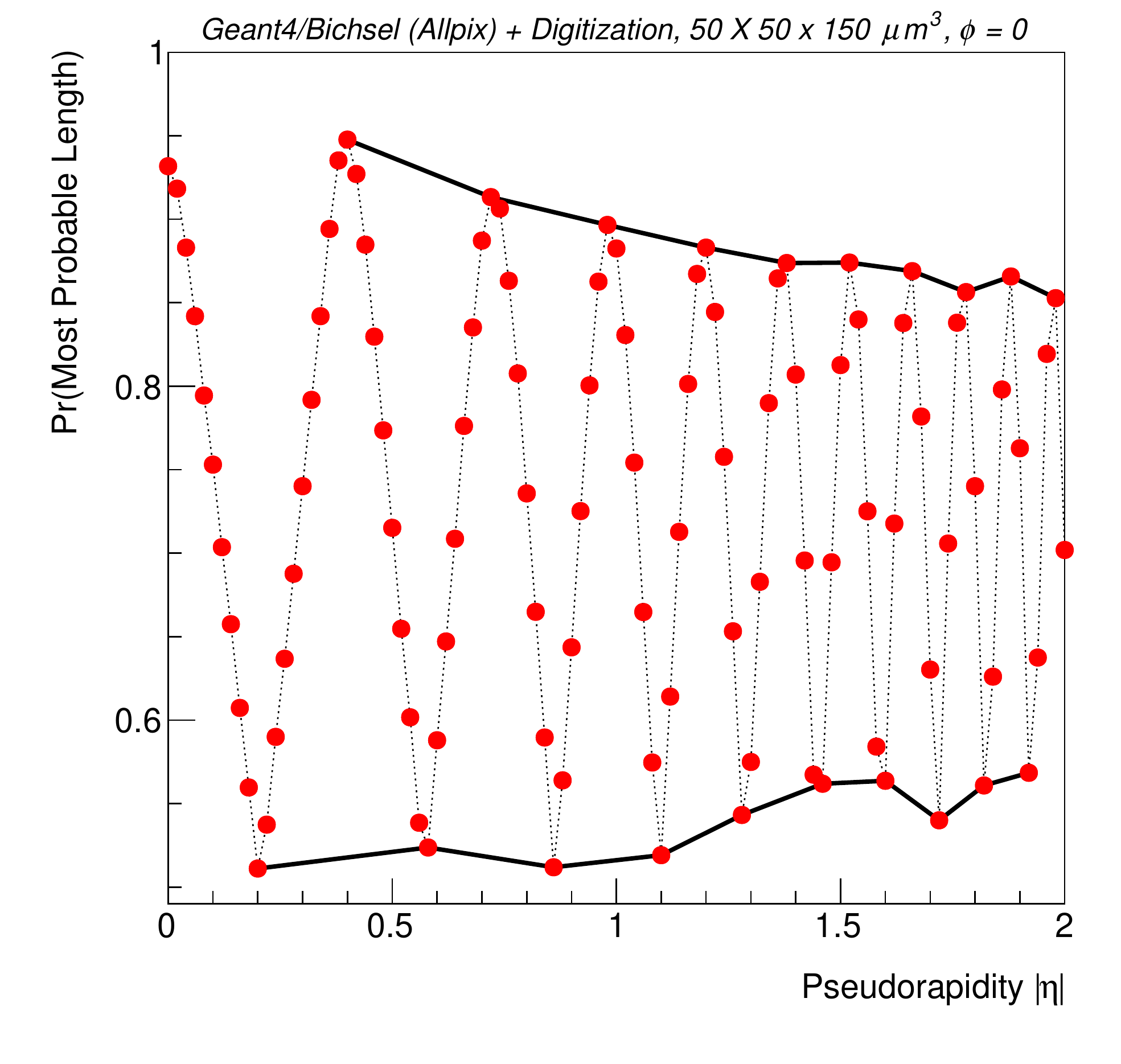}
	\caption{The probability of the most probable length, $\Pr(\text{MPL})$, as a function of $|\eta|$ for $\phi=0$ and a $50\times50\times150$ $\mu$m$^3$ pixel sensor.  The scatter seen at large $|\eta|$ is statistical, as the $\Pr(\text{MPL})$ values were determined from the simulation.  }
	\label{fig:oscill}
\end{figure*}

The periodic structure in Fig.~\ref{fig:2dprob} is a result of the pixelated geometry.  For a certain incidence angle represented by the red arrow in Fig.~\ref{fig:length}, no matter where along a pixel the particle enters the sensor, the MIP will always traverse two pixels, so will mostly give a cluster two pixels long.  Due to diffusion and threshold effects, the number of hit pixels could also be one or three with a small probability. The situation is different for an incidence angle represented by the black arrow in Fig.~\ref{fig:length}. When the particle enters to the left of the black arrow the cluster length will most likely be one pixel, while when the particle enters to the right the length will most likely be two pixels.  Therefore, the chances of $\text{length}=1$ and $\text{length}=2$ are the same, so the probability of the most probable length, $\Pr(\text{MPL})$, is near 50$\%$.  The full relationship between MPL and $|\eta|$ for $\phi=0$ is shown in Fig.~\ref{fig:oscill}.  The black arrow in Fig.~\ref{fig:length} results in the first valley in Fig.~\ref{fig:oscill}, while the red arrow in Fig.~\ref{fig:length} leads to the first peak in Fig.~\ref{fig:oscill}.  Peak and valley positions in Fig.~\ref{fig:oscill} repeat and approximately correspond to $\tan(\alpha_\text{peak})=np/d$ and $\tan(\alpha_\text{valley})=(n-0.5)p/d$, where $\alpha$ is the track angle, $n$ is any positive integer, $p$ is the pixel pitch, and $d$ is the sensor thickness.  The upper and lower bold lines in Fig.~\ref{fig:oscill} connect such peak and valley points.  

In what follows we determine the position resolutions for the peak and valley points using the cluster shape classification method described in Sec.~\ref{sec:calculation} as these resolutions provide upper and lower bounds, respectively. 
Resolutions for the $x$ and $y$ directions are analyzed separately. Since our simulation setup is symmetric, in principle $x$ and $y$ are interchangeable, but the angular ranges explored in the two directions are different. In the $x$ 
direction we only go up to $50^{\circ}$ incidence angle (where 0 is normal incidence), 
which is equivalent to $|\eta|\approx1$, while in the $y$ direction we scan up to $|\eta|=2$, which is rather shallow incidence (74.6$^{\circ}$).
As we will show, shallow incidence angles result in rather different resolutions along and transverse to the incidence direction.

\subsection{\label{sec:yres} Resolution in the $y$ direction}

\begin{figure*}[h!]
	\centering
	\subfigure[$\phi=0$]{\includegraphics[width=0.49\textwidth]{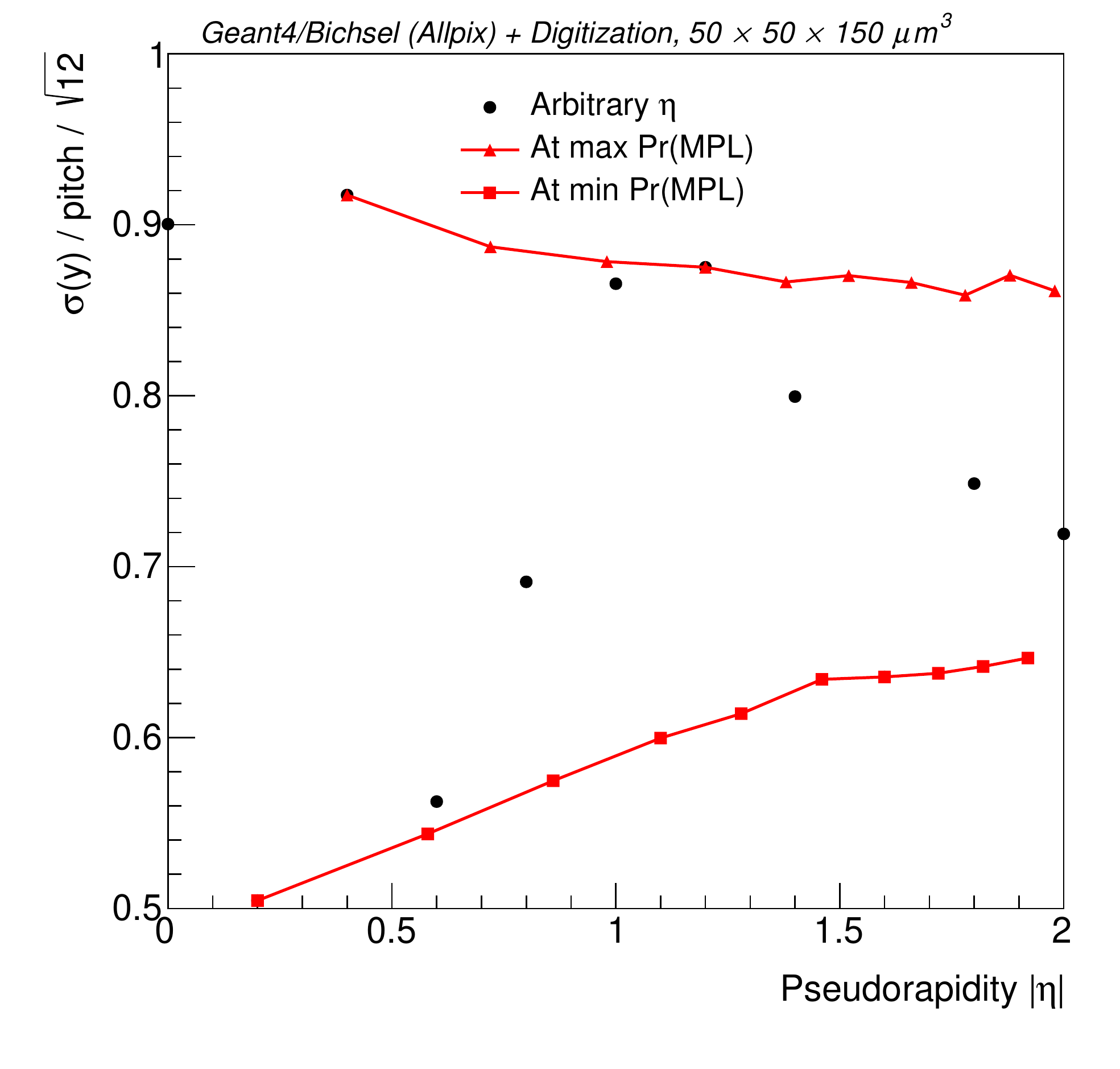}}
	\subfigure[Varying $\phi$]{\includegraphics[width=0.49\textwidth]{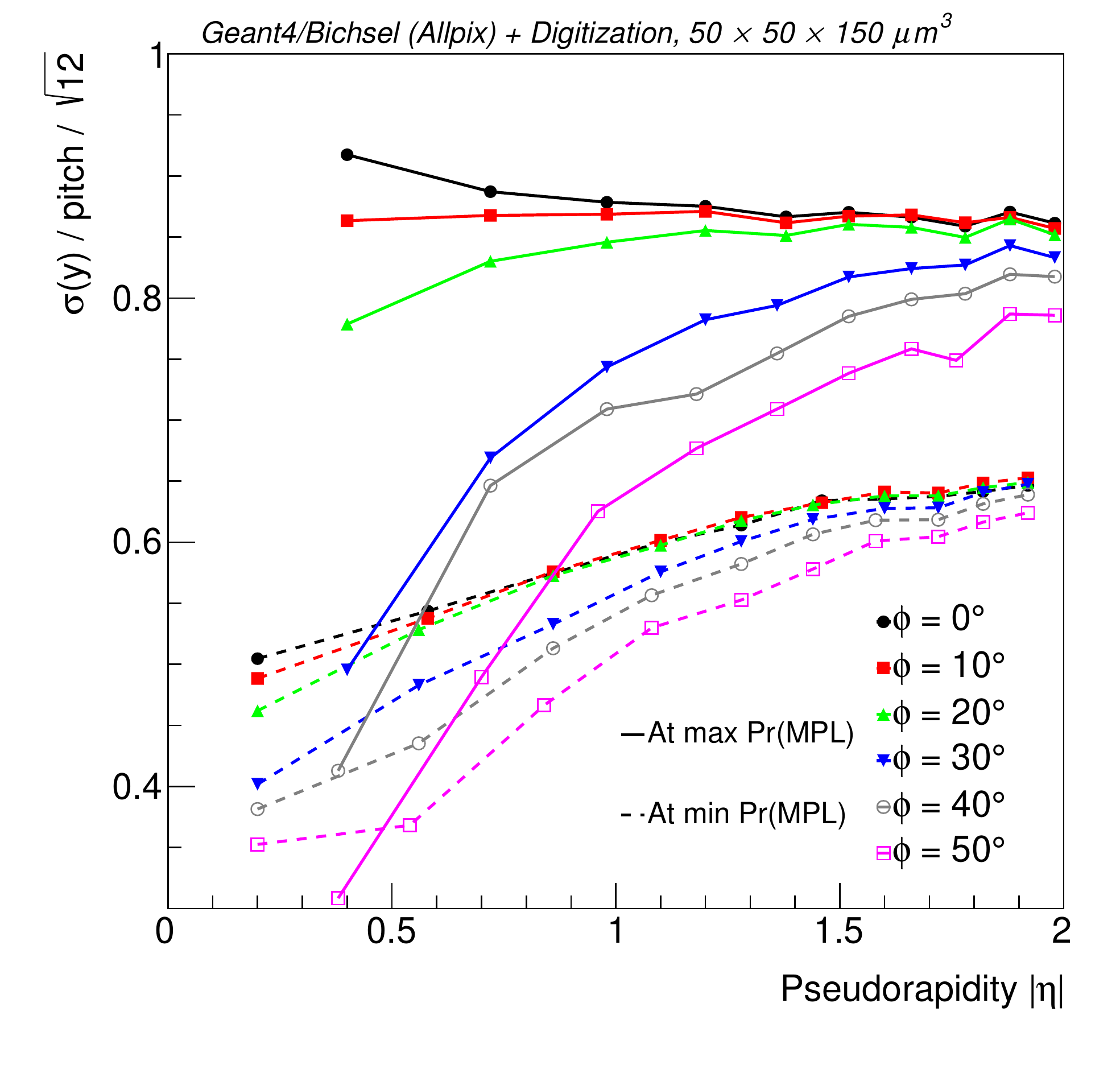}}
	\caption{The optimal $y$ position resolution normalized to pitch/$\sqrt{12}$ vs. $|\eta|$ for a $50\times 50\times 150$ $\mu$m$^3$ pixel sensor.
The envelope of peak and valley points is shown by solid/dashed lines for
(a) $\phi=0$, including a selection of arbitrary $|\eta|$ incidence angles,
and (b) for different $\phi$ values.}
	\label{fig:stdyeta}
\end{figure*}

Fig.~\ref{fig:stdyeta}(a) shows the optimal $y$ position resolution ($\sigma(y)$) normalized to pitch/$\sqrt{12}$ for $50\times50\times150$ ${\mu}$m$^3$ pixels and $\phi=0$ (normal incidence in the $x$ direction).  Points with a maximum (peak) or minimum (valley) in the probability of the most probable single cluster length are connected with red lines.  As described in the previous section, the most probable length is mostly responsible for the structure as a function of $|\eta|$ and therefore sets the resolution envelope. Points for a few arbitrary incidence angles are also shown to confirm that they lie within the envelope.  Figure~\ref{fig:stdyeta}(b) shows the resolution envelopes for different $\phi$ values (the incidence angle in the $x$ direction).  The resolution is almost always below 90\% of pitch/$\sqrt{12}$ across the entire $|\eta|$ range probed.  At small $|\eta|$ and large $\phi$, the most probable length in the $y$ direction becomes less relevant for setting the resolution boundaries because the shape in the $x$ direction dominates the entrance position.  Such extreme azimuthal incidence angles are not common in typical pixel detector layouts.  


\begin{figure*}[h!]
	\centering
	\subfigure[Large pixel size in $y$ direction]{\includegraphics[width=0.49\textwidth]{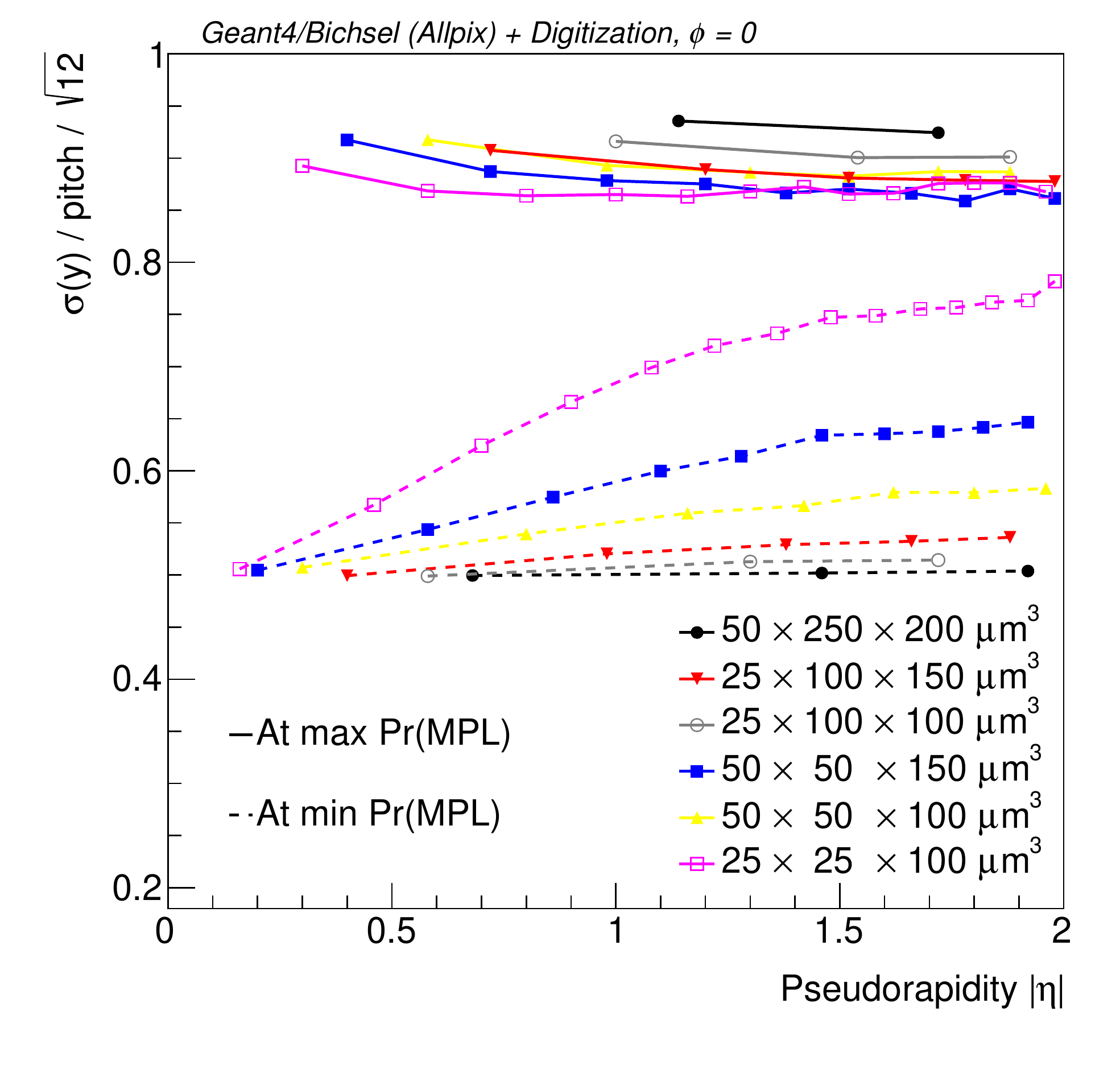}}
	\subfigure[Small pixel size in $y$ direction]{\includegraphics[width=0.49\textwidth]{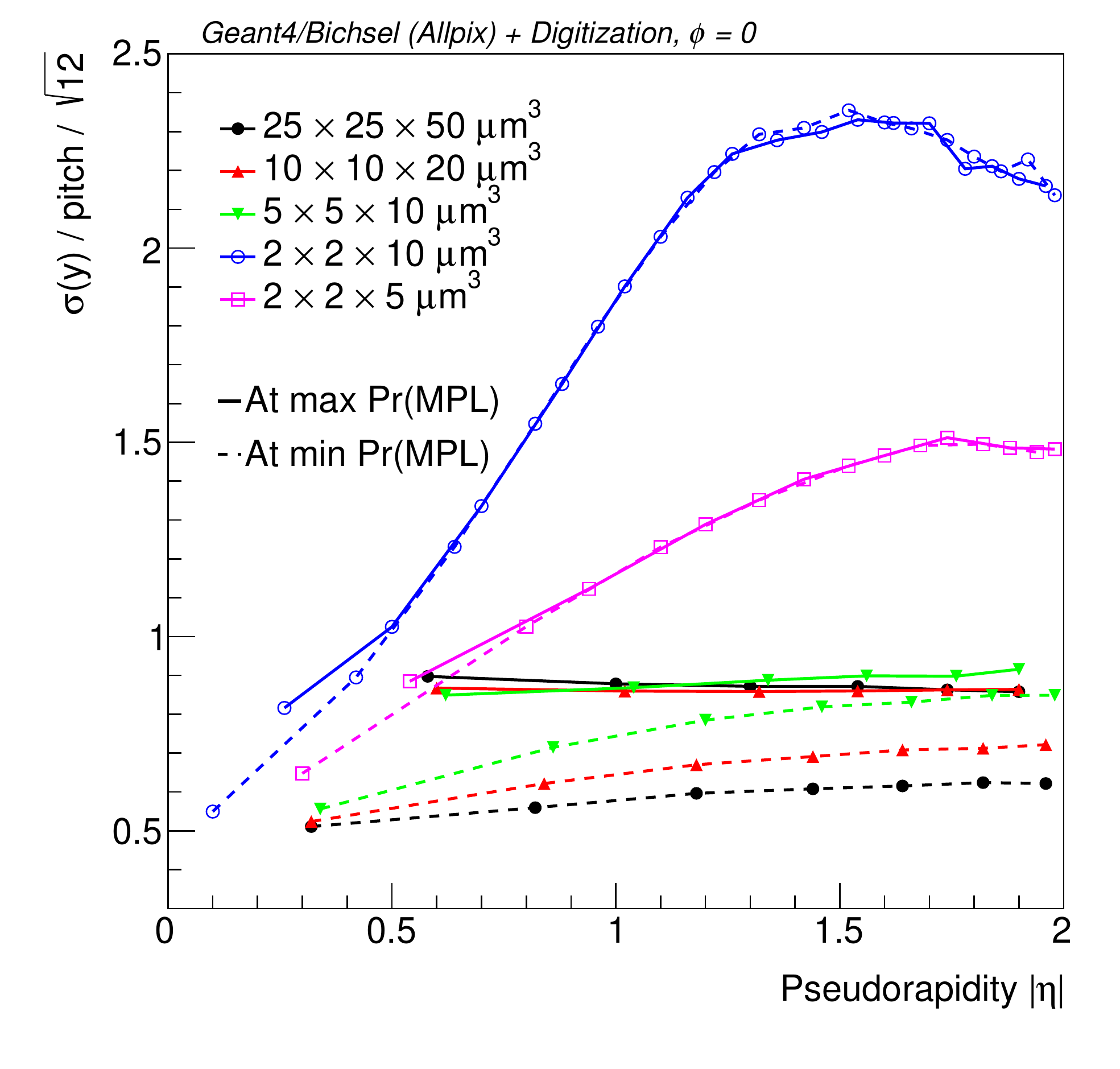}}
	\caption{The optimal $y$ position resolution normalized to pitch/$\sqrt{12}$ vs. $|\eta|$ (with $\phi=0$) for 
(a) larger pixel sizes and (b) smaller pixel sizes.
The envelope of peak and valley points is shown by solid/dashed lines.}
	\label{fig:stdyphi}
\end{figure*}

The optimal $y$ position resolutions for all the sensor geometries of Table~\ref{tab:pitch} are shown in Fig.~\ref{fig:stdyphi}, where the sensors 
have been divided into larger pixel sizes (Fig.~\ref{fig:stdyphi}(a)) and smaller pixel sizes (Fig.~\ref{fig:stdyphi}(b)). As before, the points corresponding to the peak and valley of the $\Pr(\text{MPL})$ distribution are connected with lines to illustrate resolution envelopes. 
For the larger pixels, the resolution at $\Pr(\text{MPL})$ peak (top line) has little dependence on pixel size or $|\eta|$, 
and is around 90\% of $\text{pitch}/\sqrt{12}$.
The resolution at $\Pr(\text{MPL})$ valleys, on the other hand,
depends strongly on pixel size and $|\eta|$ for square pixels, but has little dependence on size or $|\eta|$ for rectangular pixels.



The small pixel sizes (Fig.~\ref{fig:stdyphi}\,b), show a similar trend as the large ones, except for the $2\times 2\,\mu$m$^{2}$ pixels, 
which have a strikingly different behavior. For these pixels the peak and valley resolutions completely overlap
and, for the first time, the resolution becomes worse than $\text{pitch}/\sqrt{12}$ as $\eta$ increases. 
The reason is ionization statistics -- a familiar concept that limits resolution in gaseous detectors. 
In such small pixels there is a non-negligible probability for a minimum ionizing particle to 
traverse the pixel without any charge deposition. 
The distance between energy deposits for a MIP in silicon has a mean of 0.25 $\mu$m.

\subsection{Resolution in the $x$ direction}

\begin{figure}[h!]
	\centering
	\includegraphics[width=0.5\textwidth]{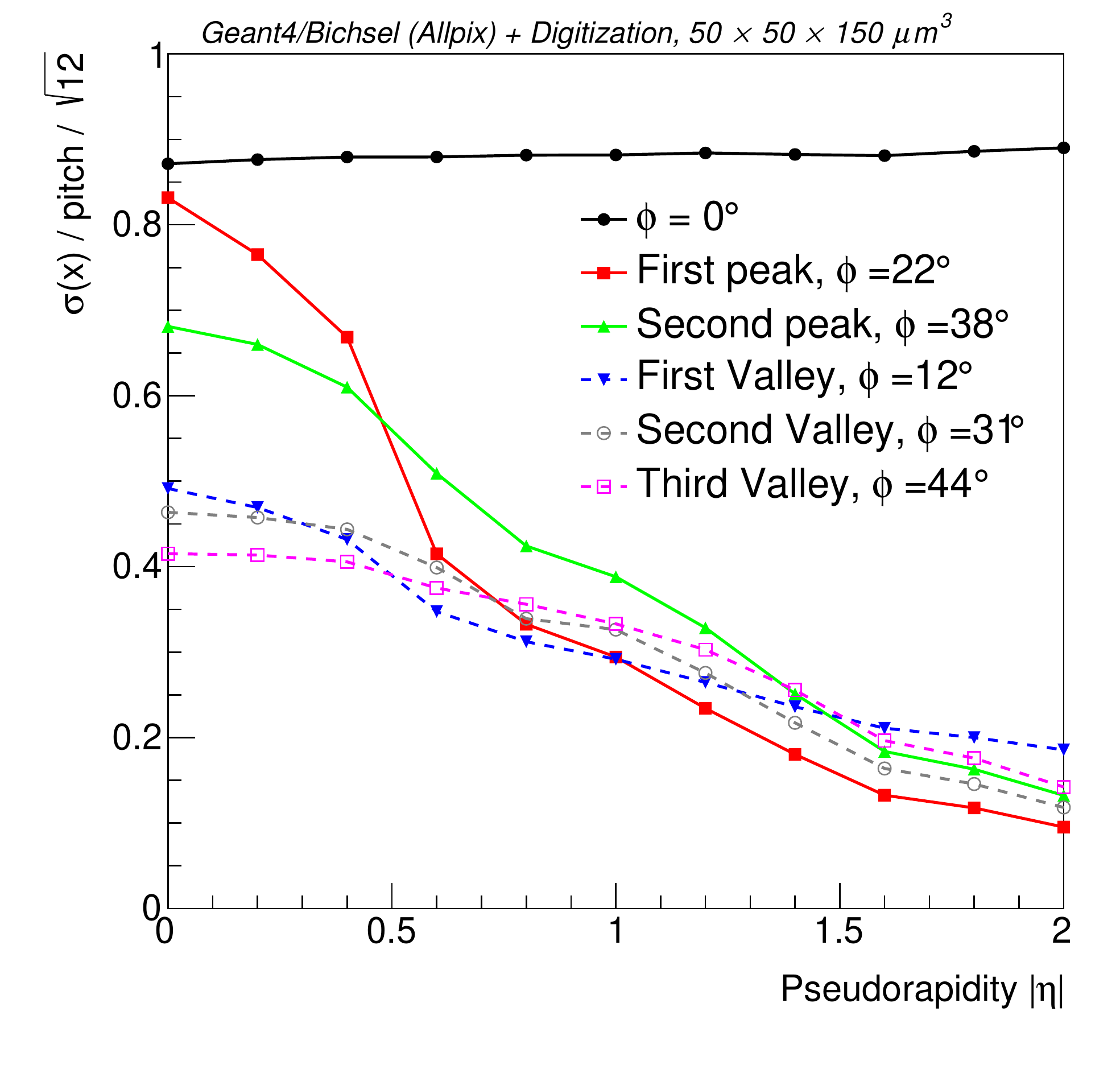}
	\caption{The optimal $x$ position resolution normalized to pitch/$\sqrt{12}$ vs. $|\eta|$ at $\phi=0$ and for various $\phi$ that correspond to the minima of $\Pr(\text{MPL})$ for length along the $x$ direction.}
	\label{fig:stdx}
\end{figure}

From the symmetry of Fig.~\ref{fig:2dprob}\,(left) it is clear that the $x$ position resolution 
vs. $\phi$ will be the same as the $y$ resolution vs. $\theta$,
which has already been covered in Sec.~\ref{sec:yres}. In this section we discuss the $x$ resolution vs. $\eta$, for a few selected $\phi$ values
lying on the peak and valley bounds when $|\eta|=0$.
Figure \ref{fig:stdx} shows the optimal $x$ position resolution normalized to pitch/$\sqrt{12}$ as a function of $|\eta|$ for a $50\times 50\times 150$ $\mu$m$^3$ sensor. 
Except for the curve at $\phi=0$, 
the other lines correspond to the minima and maxima of $\Pr(\text{MPL})$ for length along the $x$ direction.
It is clear that a detector geometry assuring a non-zero $\phi$ incidence will benefit from a significant resolution gain due to the longer cluster-size in the $y$ direction as $|\eta|$ increases. 
The optimal $x$ position resolution is better than $0.4 \times \text{pitch}/\sqrt{12}$ when $|\eta|>0.6$ and even better than $0.2 \times \text{pitch}/\sqrt{12}$ when $|\eta|>1.5$. The reason peak and valley $\phi$ lines intersect as $|\eta|$ increases is that these $\phi$ values no longer correspond exactly to probability peaks and valleys at larger $|\eta|$: 
they are the probability peak and valley points at $|\eta|=0$. At $\phi$=0 the resolution does not change with $\eta$ up to $|\eta|<2$, which corresponds to an 11 pixel long cluster.  
This is because the particle trajectory is exactly aligned with the sensor grid, 
so it maintains a constant distance to the pixel boundaries. Charge sharing only occurs when the position
is very near a pixel boundary. While a longer path length at higher $\eta$ does result in more 
charge to be shared, the charge per pixel decreases with $\eta$, reducing the sharing probability per pixel, which 
compensates the benefit of longer path. Thus the resolution remains constant when $\phi$=0. 
However, as soon as $\phi \neq 0$, the particle trajectory is more and more likely to 
cross pixel boundaries as the cluster length increases. Proximity to pixel boundaries leads to charge sharing and 
increased resolution.

%
%
%

\section{\label{sec:sharing} Impact of noise and crosstalk}

The results shown above did not include noise or electronic charge sharing (crosstalk). In this section, we investigate their impact
by adding crosstalk and Gaussian noise to selected simulations. 
We vary the threshold to noise ratio (T/N) and the fraction of charge shared electronically to each neighbor pixel ($F_{sh}$)
over a wide range. (In existing pixel detectors T/N is between 15 and 30\cite{CMST/N,ATLAST/N}, while the $F_{sh}$ is about 2\%\cite{rossi2006pixel}.)
We add crosstalk first, so that only true collected charge (not noise) 
is electronically shared. We start by reducing the collected charge, $Q_i$ in each pixel by $4F_{sh}$, and then add $Q_iF_{sh}$ to each 
neighbor (left, right, top, bottom). We finally add Gaussian noise to every pixel (hit or not). 

\begin{figure}[h!]
	\centering
	\includegraphics[width=0.5\textwidth]{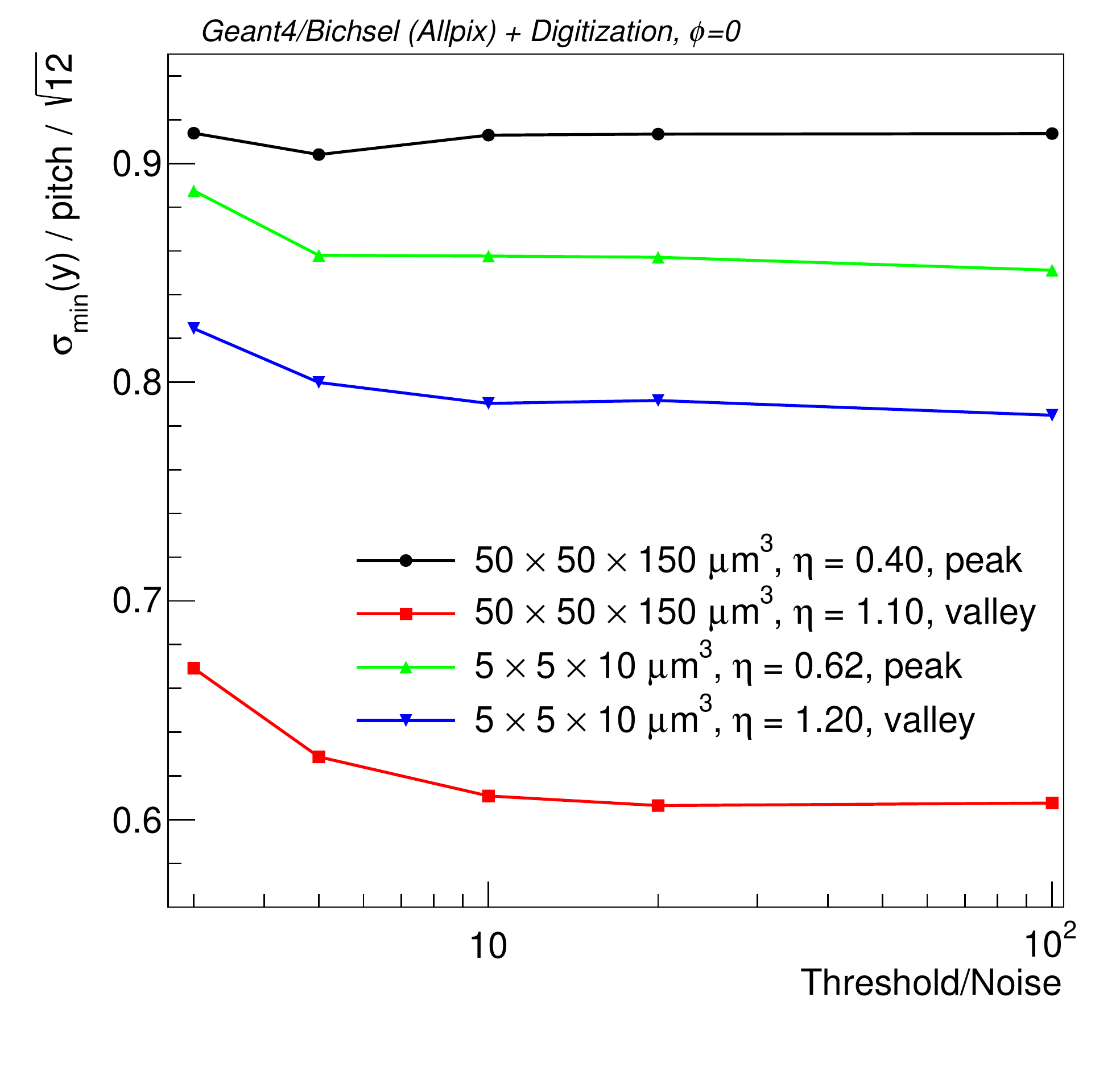}
	\caption{The $y$ position resolution normalized to pitch/$\sqrt{12}$ vs. the noise. $F_{sh}$ is 2\% for this plot. The optimal position resolution converges to the case without noise when T/N increases.}
	\label{fig:stdywithnoise}
\end{figure}

Figure \ref{fig:stdywithnoise} shows the $y$ resolution as a function of T/N for two pixel sizes, for probability peak and valley $\eta$ values.
In this case we fix $F_{sh}=2\%$. 
When T/N increases, all the lines converge to the optimal position resolution without noise. 
Degradation becomes significant only for T/N $\leq 5$, which means the results obtained earlier are accurate for typical detectors. 

\begin{figure}[h!]
	\centering
	\includegraphics[width=0.5\textwidth]{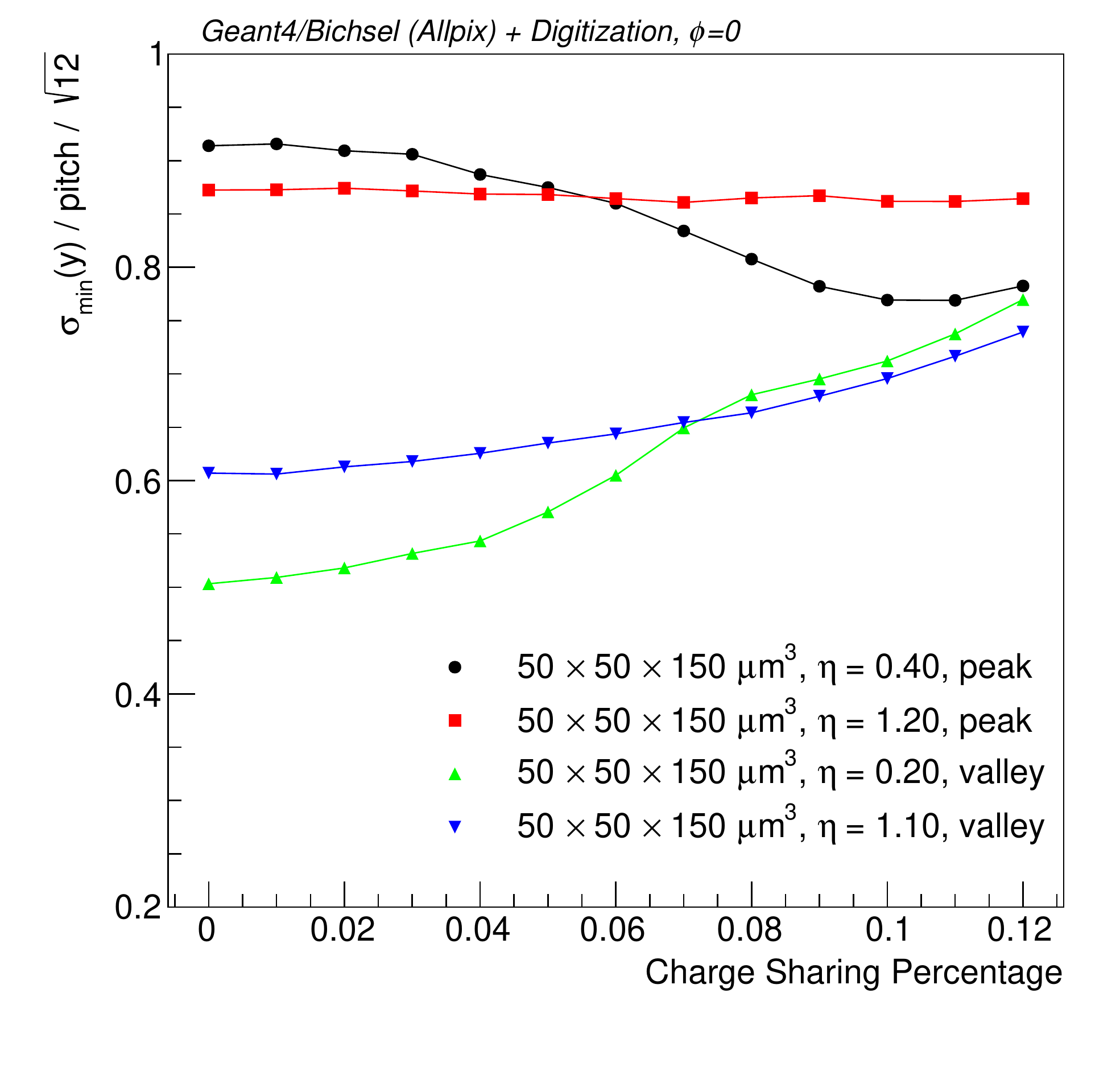}
	\caption{The $y$ position resolution normalized to pitch/$\sqrt{12}$ vs. the charge sharing percentage. The T/N for this plot is 10. Increased charge sharing is seen to degrade or improve the resolution depending on $\eta$. 
Typical detectors have $F_{sh}$ values around 2\%.}
	\label{fig:stdywithchargesharing}
\end{figure}

To investigate the effect of crosstalk we scan $F_{sh}$ from 0 to 12\% in Fig.~\ref{fig:stdywithchargesharing}, while fixing T/N=10.  
The typical 2\% crosstalk is seen to have a negligible impact.  However, a large amount of electronic charge sharing can not only degrade, 
but interestingly also improve the resolution depending on incidence angle. We believe this is because large crosstalk can act
as a dynamic threshold adjustment, effectively reducing the threshold of pixels around a hit, which makes those pixels able to detect 
smaller charge. This suggests artificially enhanced crosstalk may have potential benefits. (Exploring this concept is beyond the scope
of this paper.)  

\section{Conclusions}
\label{sec:concl}

We have presented a comprehensive study of the pixel detector position resolution obtainable with binary readout.  The results show how the resolution depends on angles and sensor geometry. Until the pixel pitch becomes so small as to be comparable to the distance between energy deposits in silicon, the resolution is always better, and in some cases much better, than the commonly assumed $\text{pitch}/\sqrt{12}$.  This information can be of practical use for charged particle track fitting and design of new detectors. This work is particularly relevant for future experiments proposing to implement very small pixels. As long as there are multiple cluster shapes produced with significant probability, binary readout can achieve excellent performance, without the need for charge information.


\section{Acknowledgments}

This work was supported by the U.S.~Department of Energy, Office of Science under contract DE-AC02-05CH11231 and by the China Scholarship Council.

\bibliographystyle{elsarticle-num}
\bibliography{myrefs.bib}{}

\end{document}